\documentclass[conference]{IEEEtran}
\IEEEoverridecommandlockouts
\usepackage{cite}
\usepackage{amsmath,amssymb,amsfonts}
\usepackage{algorithmic}
\usepackage{graphicx}
\usepackage{textcomp}
\usepackage{xcolor}
\usepackage{verbatim}
\def\BibTeX{{\rm B\kern-.05em{\sc i\kern-.025em b}\kern-.08em
    T\kern-.1667em\lower.7ex\hbox{E}\kern-.125emX}}
\begin{document}

\title{Advances in Self-Supervised Learning for Synthetic Aperture Sonar Data Processing, Classification, and Pattern Recognition\\
}

\author{\IEEEauthorblockN{Brandon W. Sheffield, IEEE Member, Frank E. Bobe III, Bradley Marchand, Matthew S. Emigh,  IEEE Member}
\IEEEauthorblockA{\textit{Naval Surface Warfare Center Panama City Division} \\
}
}

\maketitle

\begin{abstract}
Synthetic Aperture Sonar (SAS) imaging has become a crucial technology for underwater exploration because of its unique ability to maintain resolution at increasing ranges, a characteristic absent in conventional sonar techniques. However, the effective application of deep learning to SAS data processing is often limited due to the scarcity of labeled data. To address this challenge, this paper proposes MoCo-SAS that leverages self-supervised learning (SSL) for SAS data processing, classification, and pattern recognition. The experimental results demonstrate that MoCo-SAS significantly outperforms traditional supervised learning methods, as evidenced by significant improvements observed in terms of the F1-score. These findings highlight the potential of SSL in advancing the state-of-the-art in SAS data processing, offering promising avenues for enhanced underwater object detection and classification.
\end{abstract}


\section{Introduction}


Acoustic sonar, particularly Synthetic Aperture Sonar (SAS), has emerged as a vital technology in the domain of underwater imaging, primarily due to the unique challenges that marine environments pose to traditional computer vision techniques. The often unpredictable and adverse conditions of underwater settings, characterized by poor lighting, turbidity, and suspended sediment, have made it arduous for conventional computer vision methodologies, which predominantly rely on optical camera imagery, to adapt effectively. Unlike side scan sonar (SSS) or real aperture sonar (RAS), where resolution worsens with range, SAS overcomes this limitation by coherently combining consecutive sonar pings to synthetically lengthen the aperture, resulting in constant resolution throughout the image\cite{bell1997simulation, cook2004auv, piper2002detection, 5191242, sammelmann1997high}.

In recent years, Deep Neural Networks (DNNs) have found favor over classical machine learning methodologies due to their proficiency in autonomously discovering features in data, thereby eliminating the need for manual feature crafting by subject matter experts. However, a notable limitation associated with DNNs is their dependence on substantial volumes of labeled data and considerable computational power to learn robust features. In the context of SAS, the scarcity of labeled data exacerbates this problem, making it difficult to adapt DNNs to underwater imaging. Emerging as a promising solution to this predicament is Self-Supervised Learning (SSL). The growing availability of computational power and data has propelled SSL into prominence, offering a methodology that allows models to learn features in data without the need for labels. Despite its potential, the application of SSL to SAS data processing, classification, and pattern recognition has received limited attention.


This work is an iterative improvement on a previous work that demonstrated that two popular SSL approaches in SAS data produced positive results for low-labeled scenarios and the models were fine-tuned only in the last layer\cite{sheffield2023selfsupervised}. The aim of this paper is to demonstrate the potential of SSL when limited SAS-labeled data exist, by proposing the MoCo-SAS framework. The SSL model is trained on real-world SAS data to learn useful feature representations, and once the model is pretrained on unlabeled data, it serves as a feature extractor for downstream binary image classification using a Support Vector Machine(SVM). The performance of this SSL model is further evaluated and compared to its supervised learning counterpart.

\section{Background}


Early work involving unsupervised learning in SAS data processing utilized Latent Dirichlet Allocations (LDA) and Auto-Encoders (AE) to extract features of objects in SAS images unsupervised\cite{isaacs2014representational}.

Pre-training is another important aspect of SAS data processing. Multiple works have shown that the use of a pre-trained CNN for feature extraction in combination with a Support Vector Machine for classification improves classification performance \cite{huang2006large, zhu2017deep, mckay2017s, rutledge2018intelligent}. In the work of Williams \cite{williams2018convolutional, williams2019transfer}, the author explored the use of transfer learning with SAS image convolutional neural networks for improved classification of underwater objects.


The paper titled "Self-supervised Learning for Sonar Image Classification"\cite{preciado2022self} by Preciado-Grijalva et al. discusses how SSL can be a powerful approach to learn forward-look sonar (FLS) image representations without the need for large labeled data sets. The paper investigates the potential of three SSL methods (RotNet, Denoising Autoencoders, and Jigsaw) to learn high-quality sonar image representation without the need for human labels.

\section{MoCo-SAS}

\begin{figure}[h]
\centering
\includegraphics[width=0.4\textwidth]{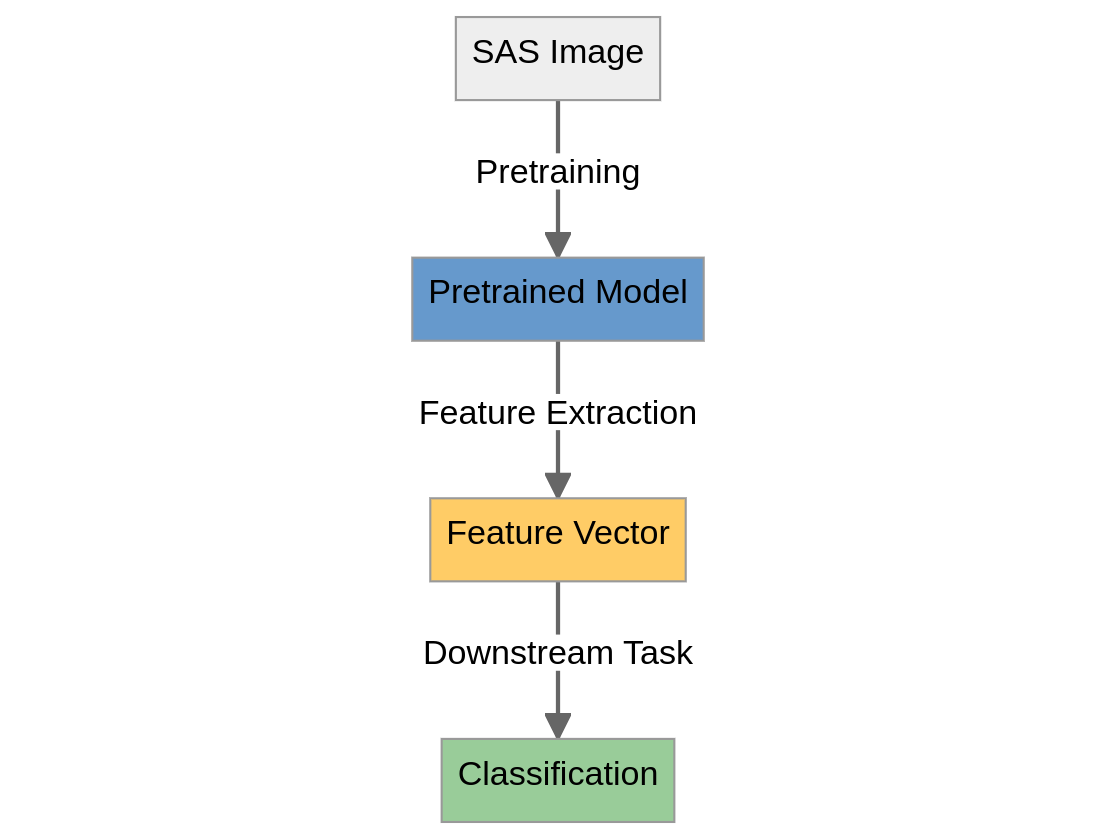}
\caption{Block diagram demonstrating the MoCo-SAS classification processing chain. The process begins with the pretraining phase, where the model learns to generate informative features from the unlabeled SAS data. These features are then passed through a feature extractor, which transforms the raw SAS data into high-level representations. These representations are then used in the downstream tasks, such as classification using an SVM classifier.}
\label{fig:ssl_framework}
\end{figure}


The MoCo-SAS framework is built upon the Momentum Contrast (MoCo) approach \cite{he2020momentum}\cite{Chen2020ImprovedBW}, which is a strategy for contrastive loss learning. To enhance the model's ability to learn robust features from unlabeled data during the pretraining phase, we have integrated a data augmentation module specifically tailored for Synthetic Aperture Sonar (SAS). MoCo was selected as the foundation for this framework primarily due to its inclusion of a queue or 'memory bank.' This queue stores numerous negative samples from previous batches, thereby mitigating the substantial computational expense associated with the need for large training batches when calculating the contrastive loss.


MoCo-SAS is composed of a backbone network based on the ResNet\cite{he2016deep} architecture and a multi-layer perceptron projection head. The backbone network extracts low-level features from the input data, while the projection head maps these features to a latent space where contrastive learning is performed. In this study, a custom version of the ResNet18 architecture shown in Table \ref{tab:custom_resnet}, is customized to handle 1-channel SAS imagery that is used for both the MoCo-SAS backbone as well as a comparative version of supervised learning in model evaluation. The final classification head of the standard ResNet18 is removed, making the model suitable for downstream tasks. After pretraining, the model acts as a feature extractor, transforming raw SAS data into high-level representations that serve as input for an SVM classifier.


\begin{table*}[h]
    \centering
    \footnotesize
    \caption{Custom ResNet-18 Architecture}
    \begin{tabular}{|c|c|c|c|}
        \hline
        \textbf{Layer (type)} & \textbf{Output Size} & \textbf{Number of Parameters} & \textbf{Configuration} \\ [0.5ex]
        \hline\hline
        Conv2d & 64x112x112 & 3200 & Kernel = 7, Stride = 2, Padding = 3 \\
        \hline
        BatchNorm2d & 64x112x112 & 128 & Eps = 1e-05, Momentum = 0.1, Affine = True \\
        \hline
        ReLU & 64x112x112 & 0 & Inplace = True \\
        \hline
        MaxPool2d & 64x56x56 & 0 & Kernel = 3, Stride = 2, Padding = 1, Dilation = 1 \\
        \hline
        ... & ... & ... & ... \\
        \hline
        AdaptiveAvgPool2d & 512x1x1 & 0 & Output Size = 1 \\
        \hline
    \end{tabular}
    \label{tab:custom_resnet}
\end{table*}

\section{Experimental Setup}

The MoCo-SAS model is first pre-trained on a large unlabeled SAS data set using the PyTorch Lightning framework \cite{Falcon_PyTorch_Lightning_2019}. The backbone network of the MoCo-SAS model is a ResNet18 architecture.

An SVM from scikit-learn\cite{scikit-learn} is trained using the high-level features extracted from the pretrained MoCo-SAS model. Training of this classifier and the comparative ResNet18 supervised learning model is carried out using a fraction of the labeled SAS data set, where the fraction is a variable parameter ranging from 1\% to 100\% of the data set. The performance of the MoCo-SAS framework is evaluated based on the F1-score of the SVM classification model in the test data set and compared against a typical supervised learning ResNet18 classification model that is initialized with random weights. The comparison between MoCo-SAS and ResNet18 seeks to show that initialization of pre-trained weights from unlabeled data is beneficial for improving classification performance when there are limited labeled data.

The experiments are carried out using the Weights and Biases (WandB)\cite{wandb} tool for the tracking of the experiment. The results of each run, including the run configuration and the performance metrics, are logged to WandB. The best run is determined based on the highest F1-score achieved across all runs on the test data set.

\subsection{Data set}


The high-resolution nature of Synthetic Aperture Sonar (SAS) data often exceeds the processing capabilities of contemporary GPUs. This requires the division of the imagery into smaller, manageable sections, often referred to as tiles, snippets, or chips. To ease the burden of creating a data set comprising snippets that contain objects on the sea floor, we employ the FRED anomaly detector\cite{brandes2021environmentally} which is a fusion of two simple and efficient detection algorithms: the raw energy detector and the KL divergence Reed-Xialo (RX) \cite{reed1990adaptive}. Each snippet, both from the low-frequency (LF) and high-frequency (HF) ranges, is resized to a dimension of 224x224. These snippets are then stacked to form a multiband SAS image with dimensions 2x224x224. The primary experiments conducted in this study focus on the HF portion of the images, and a later ablation study considers both single and dual frequency channels.

\begin{table}[h]
\centering
\begin{tabular}{|c|c|}
\hline
\textbf{Data set} & \textbf{Number of Samples} \\ \hline
Pretrain & 400,000 \\ \hline
Train & 6,000 \\ \hline
Test & 2,000 \\ \hline
\end{tabular}
\caption{The pretrain data set is unlabeled while the train and test data sets are labeled. Both the train and test data sets are balanced with positive and negative instances.}
\label{tab:dataset_statistics}
\end{table}

\subsection{Augmentations}

This study applied a multi-view data augmentation strategy to SAS imagery for pretraining the MoCo-SAS model. This strategy is crucial for learning robust representations, as it encourages the model to identify and focus on the invariant features of the same image under different transformations. However, in the downstream training and test phases, only the horizontal flip augmentation is applied.

\begin{figure}[h]
\centering
\includegraphics[width=0.4\textwidth]{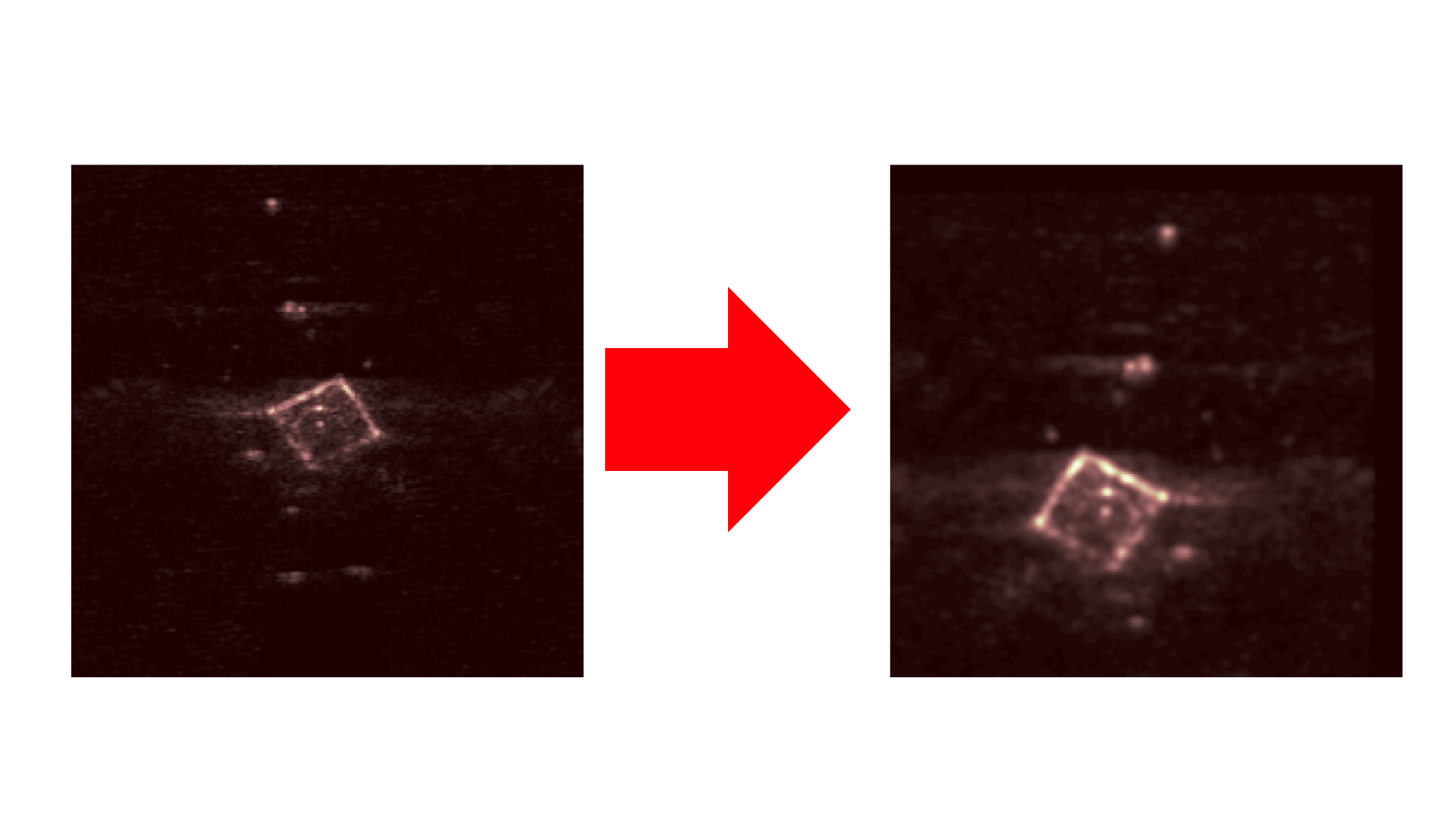}
\caption{Before and after transformation of a SAS image with a crab trap that has data augmentations applied. Augmentations must be carefully selected such that the semantic content of the image remains the same.}
\label{fig:sas_augs}
\end{figure}

The multiview data augmentation process generates two distinct views from the same input image. Each view is produced by applying a probabilistic set of transformations, including random horizontal flip, random resized crop, subtle random rotation, random affine transformations, random Gaussian blur, and speckle noise as can be seen in Figure \ref{fig:sas_augs}. The latter introduces a type of interference commonly found in radar systems, including SAS, thereby enhancing the model's resilience to such noise in real-world scenarios.

\subsection{Pretraining}

The MoCo-SAS model is pre-trained on the pre-trained data set using the hyperparameters in Table \ref{tab:hyperparameter_table}. During pretraining, each mini-batch comprises two views for each image, produced by the aforementioned data augmentation process. The model learns to associate the two views of the same image (a positive pair) while contrasting them from all other views (negative pairs) in the minibatch. By employing this multiview data augmentation strategy, the model learns to recognize and understand the underlying invariant features of the SAS imagery, thereby achieving better performance in downstream tasks.

\begin{figure}[h]
\centering
\includegraphics[width=0.4\textwidth]{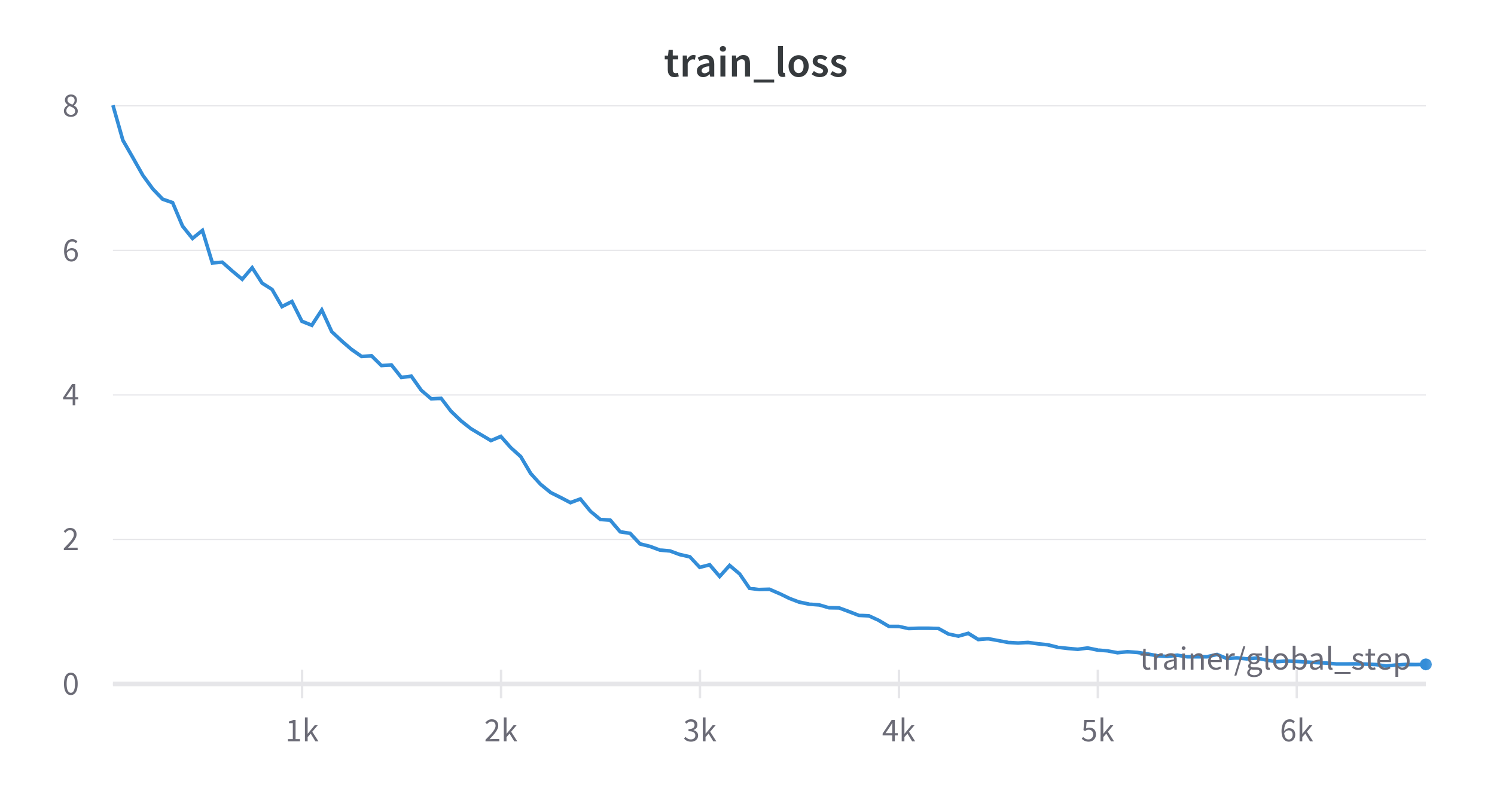}
\caption{Loss curve of the pretraining phase. The x-axis represents the epochs, and the y-axis represents the loss.}
\label{fig:loss_curve}
\end{figure}

This process is repeated for multiple epochs, with the model's parameters being updated after each batch to minimize the contrastive loss. To prevent overfitting and ensure efficient learning, early stopping is applied during training. This stops the training process if there is no significant decrease in loss for 10 consecutive epochs. With 8 Nvidia A6000 GPUs, training halted at 6 hours and 39 minutes at 184 epochs as shown in Figure \ref{fig:loss_curve}.

\begin{table}[h]
\centering
\begin{tabular}{|c|c|}
\hline
\textbf{Hyperparameters} & \textbf{Values}  \\ \hline
Backbone                & ResNet18      \\ \hline
Channels                & 1     \\ \hline
Batch Size              & 1536          \\ \hline
Optimizer               & AdamW   \\ \hline
Scheduler               & One Cycle LR   \\ \hline
Loss                    & NTXent      \\ \hline
Learning Rate           & 0.3      \\ \hline
Weight Decay            & 0.0001    \\ \hline
Memory Bank             & 4096         \\ \hline
Temperature             & 0.07        \\ \hline
Momentum                & 0.99      \\ \hline
Warmup Epochs           & 10        \\\hline
\end{tabular}
\caption{MoCo-SAS hyperparameters that are used in the pretraining stage.}
\label{tab:hyperparameter_table}
\end{table}

\subsection{Metrics}


The effectiveness of the proposed MoCo-SAS framework is assessed using three widely applied evaluation criteria: Precision, Recall, and F1-score. These metrics provide a comprehensive evaluation of the model's performance, quantifying its ability to accurately identify positive instances (objects) and its capacity to limit false alarms (misclassifying clutter or negative instances as positive).

Precision is defined as the proportion of true positives (TP) out of all predicted positives. It measures the classification model's ability to return only relevant instances. Specifically, it calculates the fraction of instances correctly identified as objects out of all instances that the model classified as such.

$$\text{Precision} = \frac{\text{True Positives}}{\text{True Positives} + \text{False Positives}}$$

Recall, also known as Sensitivity or True Positive Rate, is the ratio of true positives (TP) to all actual positives. Measures the ability of the classification model to identify all relevant instances. Specifically, it calculates the fraction of actual objects that the model correctly identified.

$$\text{Recall} = \frac{\text{True Positives}}{\text{True Positives} + \text{False Negatives}}$$

F1-score is the harmonic mean of Precision and Recall, providing a balanced representation of these two metrics.

$$\text{F1-score} = 2 \times \frac{\text{Precision} \times \text{Recall}}{\text{Precision} + \text{Recall}}$$

In the experiments, the F1-score of the classification model is evaluated on the test data set, as it provides a balanced measure of its performance in terms of both Precision and Recall. In the context of the SAS object detection task, both missing an object (a false negative) and falsely identifying clutter as an object of interest (a false positive) can carry substantial implications.

\section{Results and Discussion}


To validate the effectiveness of the approach, extensive experiments are conducted using real-world SAS data sets containing various types of underwater objects. The performance of the MoCo-SAS framework is compared with a traditional supervised learning ResNet18 model for SAS data processing, classification, and pattern recognition.

\begin{figure}[h]
\centering
\includegraphics[width=0.4\textwidth]{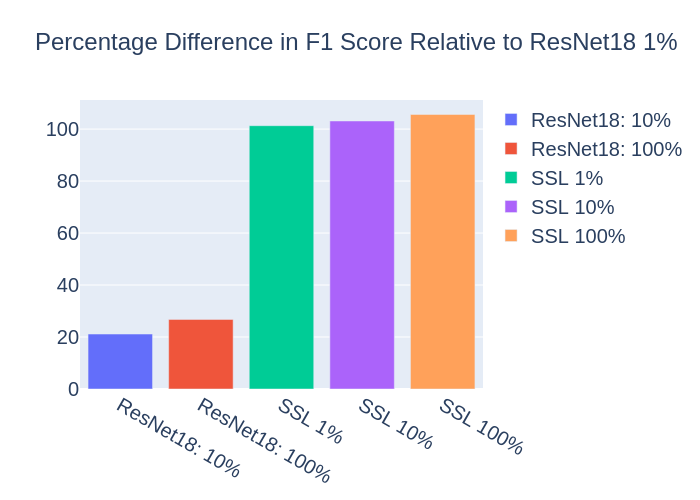}
\caption{Comparison of F1-score on the test data set using the proposed MoCo-SAS, denoted as SSL, approach versus the typical ResNet18.}
\label{fig:f1_scores}
\end{figure}

In extensive experiments, the proposed MoCo-SAS framework demonstrated significant improvements over traditional supervised learning methods in terms of F1-score. Figure \ref{fig:f1_scores} summarizes the results of the experiments, indicating the F1-score for each method in the test data set. 


The MoCo-SAS approach outperforms the ResNet18 supervised learning method by significant margins in terms of F1-score. This is primarily due to the framework's ability to leverage unlabeled data to learn useful representations, thereby overcoming the scarcity of labeled data in SAS applications. Moreover, the robustness of the learned features is evident from the consistent performance of the method across various underwater environments. These features capture the underlying patterns in the SAS data, allowing the classification model to generalize well to new data. These results confirm the effectiveness of the MoCo-SAS framework for SAS data processing. By exploiting the structure of unlabeled data, the method is capable of learning robust representations and achieving superior performance in object detection tasks.

\subsection{Discussion on SVM Performance}

The results indicate that the Support Vector Machine (SVM) classifier outperforms other classifiers for downstream tasks in the MoCo-SAS framework. One possible explanation for this superior performance is the effectiveness of the SVM in high-dimensional spaces. Given that the data features extracted from the pre-trained model are high-dimensional, SVM is likely to perform well.

Additionally, SVM is robust against overfitting, particularly in situations where the number of dimensions exceeds the number of samples. This characteristic may have contributed to its success in the task. Furthermore, the use of a subset of training points in the decision function makes SVM memory efficient, which can be beneficial for large-scale data.

\subsection{Ablation Study Results}

Table \ref{tab:ablation_depth} presents an ablation study on the impact of backbone depth (ResNet18, ResNet34, ResNet50), label percentages, and the number of channels on the performance of a SVM classifier. From the results, it is evident that the depth of the backbone, the percentages of labels and the number of channels all have a significant impact on the performance of the SVM classifier.

\begin{table}[h]
\begin{tabular}{|c|c|c|c|c|}
\hline
\textbf{Rank} & \textbf{Backbone} & \textbf{Label \%} & \textbf{Channels} & \textbf{Performance Tier} \\ \hline
1 & ResNet50 & 100 & 2 & High \\ \hline
2 & ResNet18 & 100 & 2 & High \\ \hline
3 & ResNet18 & 100 & 1 & High \\ \hline
4 & ResNet50 & 100 & 1 & High \\ \hline
5 & ResNet50 & 1 & 1 & High \\ \hline
6 & ResNet34 & 100 & 2 & High \\ \hline
7 & ResNet34 & 100 & 1 & High \\ \hline
8 & ResNet50 & 10 & 2 & High \\ \hline
9 & ResNet50 & 5 & 2 & Medium \\ \hline
10 & ResNet34 & 1 & 1 & Medium \\ \hline
11 & ResNet50 & 5 & 1 & Medium \\ \hline
12 & ResNet50 & 10 & 1 & Medium \\ \hline
13 & ResNet18 & 10 & 1 & Medium \\ \hline
14 & ResNet18 & 5 & 1 & Medium \\ \hline
15 & ResNet34 & 10 & 1 & Medium \\ \hline
16 & ResNet34 & 5 & 1 & Medium \\ \hline
17 & ResNet18 & 10 & 2 & Low \\ \hline
18 & ResNet18 & 5 & 2 & Low \\ \hline
19 & ResNet18 & 1 & 1 & Low \\ \hline
20 & ResNet34 & 10 & 2 & Low \\ \hline
21 & ResNet34 & 5 & 2 & Low \\ \hline
22 & ResNet50 & 1 & 2 & Low \\ \hline
23 & ResNet18 & 1 & 2 & Low \\ \hline
24 & ResNet34 & 1 & 2 & Low \\ \hline
\end{tabular}
\caption{Experiments performed to examine whether backbone model depth in MoCo-SAS improved performance for single and multi-channel scenarios with varying label percentages.}
\label{tab:ablation_depth}
\end{table}

\begin{itemize}
\item \textbf{Backbone Depth:} The deeper models (ResNet34 and ResNet50) generally outperform the shallower model (ResNet18), especially when the label percentage is high (100\%). This suggests that deeper models are capable of extracting more complex features from the data, which can lead to improved performance. However, when the percentage of labels is low (1\%), the performance of all models decreases, with ResNet34 and ResNet50 performing only slightly better than ResNet18. This could be because fewer labels provide the models with less information, making it harder to extract useful features.

\item \textbf{Label Percentages:} The performance of all models improves as the percentage of labels increases. This is expected, as more labels provide more information from which models can learn. However, even at low label percentages (1\% and 5\%), the models are still able to achieve medium to high performance, suggesting that they are able to effectively leverage unlabeled data.

\item \textbf{Number of Channels:} The number of channels also has a significant impact on performance. In general, the models perform better with two channels than with one. This could be because two channels provide more information than one, allowing the models to extract more complex features. However, the performance gain from using two channels is more pronounced for the deeper models (ResNet34 and ResNet50) than for the shallower model (ResNet18), suggesting that the deeper models are better able to leverage additional information.
\end{itemize}

These findings highlight the importance of considering the depth of the backbone, the percentages of labels, and the number of channels when training an SVM classifier for sonar imagery. Future work could explore these factors in more depth, as well as investigate other potential influences on classifier performance.

\section{Conclusion and Future Work}


This study introduced MoCo-SAS, demonstrating the potential of such techniques in advancing SAS data processing, classification, and pattern recognition. The approach leveraged contrastive learning and clustering-based methods to exploit the structure of unlabeled data and extract high-level features, contributing to the performance enhancement of the downstream tasks in image classification and object recognition. Significant improvements in the F1-score and robustness of the learned features, which highlights the generalizability of the approach.



Future research directions include exploring advanced self-supervised learning algorithms that can further enhance the learning of robust features. Integrating the framework with other sonar modalities can lead to a more comprehensive and robust system. Synthetic data augmentation and its effects on classification performance also deserve further investigation. We also intend to develop real-time processing capabilities for practical deployment in Autonomous Underwater Vehicle (AUV) object detection applications. This will involve optimizing the computational efficiency of the method to meet the real-time processing requirements of such applications.

\bibliographystyle{IEEEtran}
\bibliography{references}

\end{document}